# *In situ* studies of evolution of microstructure with temperature in heavily deformed Ti-modified austenitic stainless steel by X-ray Diffraction technique

**P.S.Chowdhury, N. Gayathri, P. Mukherjee, M. Bhattacharya, A. Chatterjee, A.Dutta[+] and P. Barat**

Variable Energy Cyclotron Centre, 1/AF Bidhannagar, Kolkata- 700064, INDIA

[+] Department of Metallurgical and Materials Engineering, Jadavpur University, Kolkata - 700032, INDIA

**ABSTRACT:**

The mechanism of the evolution of the deformed microstructure at the earliest stage of annealing where the existence of the lowest length scale substructure paves the way to the formation of the so-called subgrains, has been studied for the first time. The study has been performed at high temperature on heavily deformed Ti-modified austenitic stainless steel using X-ray diffraction technique. Significant changes were observed in the values of the domain size, both with time and temperature. Two different types of mechanism have been proposed to be involved during the microstructural evolution at the earliest stages of annealing. The nature of the growth of domains with time at different temperatures has been modelled using these mechanisms. High-resolution transmission electron microscopy has been used to view the microstructure of the deformed and annealed sample and the results have been corroborated successfully with those found from the X-ray diffraction techniques.

**1. INTRODUCTION**

Microstructures arising out due to deformation in solid polycrystalline materials are quite complex and rich in detail, encompassing many more features like texture, grain orientation and rotation than just the microstructure formed due to the generation and entanglement of dislocations. Annealing of these deformed materials results in a change of the microstructure owing to the change in the configurations of dislocation but at the same time the process involves a change in the texture and other microstructural parameters. Extensive early work on the recovery of deformed materials has been reviewed by Beck [1], Bever [2], Titchener and Bever [3]. However, during the later years, the search for quantitative physically-based models for the annealing process has shown renewed interest [4]. Various stages in the recovery during annealing are the formation of dislocation tangles and cells, annihilations of dislocation within cells, subgrain formation and subgrain growth [5]. The study of the evolution of the microstructural features during disentanglement of the dislocation network in the very early stages of annealing has not been addressed so far. This study can be carried out during annealing of an aggregate of powder sample in heavily deformed state which has high density of dislocations in the form of tangles. Moreover, the information of the preferred orientation in the powder sample is eliminated due to the randomization of the crystallographic orientation.

There are some evidences of subgrain formation and rotation during the last stages of recovery process where thin foils of few alloy systems were studied in situ at high temperature using High Voltage Electron Microscope [6, 7]. Studies of subgrain growth in polycrystals have also been carried out using Molecular Dynamics simulation [8, 9]. The sizes of the subgrains in these studies were found to vary within a range of $0.1 \mu m$ to few microns. But in the very early stages of microstructural evolution during annealing, the length scales of the substructures are much less than these subgrains. Humphrey *et al.* [10] have emphasised that the understanding of the mechanism of the formation of the subgrains from the lower length scale substructures is difficult or rather impossible as the microstructural evolution during this stage cannot be observed *in situ*. The overlap of the strain field due to the presence of high density dislocation tangles in deformed sample imposes a limitation to view the region by Transmission Electron Microscope in the very early stages of annealing. On the contrary, X-ray diffraction (XRD) technique can be a unique characterisation tool to obtain the statistically averaged information of the evolution of the spatially heterogeneous substructure of the deformed polycrystalline materials *in situ* during annealing as the wavelength of X-ray is comparable with the length scale of these dislocation substructures. For the first time we have used XRD technique to study the microstructural changes during the very early stage of annealing. In the following text these substructures will be referred to, as domain.

There are many theoretical and experimental studies involving the kinetics of the subgrain growth during the recovery period of the annealing process [11-20]. Sandström *et. al.* [21] have proposed a model describing the subgrain growth to occur in different ways: by migration of sub-boundaries or dissolution of sub-boundaries. The dynamical process involved during the earliest stage of annealing leading to the formation of domains of size around 10 nm from loose dislocation tangles has not been addressed so far due to the experimental limitation to view the domains of such length scale. Moreover the process of rearrangement of dislocations being very fast, it was not feasible to capture the evolution of the domain in the earliest stage of annealing under TEM for pure metals or very dilute alloys. The process of rearrangement of dislocations can be studied at this stage only if the kinetics can be slowed down by addition of alloying elements. There are substantial evidences that fine particle dispersion may exert strong pinning effect on the growth of subgrains [22, 23].

We have carried out our studies on Ti-modified austenitic stainless steel (D9), rich in alloying elements, which is an important structural material for Prototype Fast Breeder Reactors [24]. This material shows a very good combination of high temperature tensile and creep strength properties, irradiation creep resistance and resistance to irradiation induced void swelling [25, 26]. We have extracted the powder sample from the bulk annealed material by introducing severe deformation using fine jewellery file. The evolution of the microstructural changes as a function of temperature and time of the powdered sample has been studied *in situ* with the help of high temperature X-ray diffractometer. We have studied the variation of the domain size and microstrain using X-ray diffraction Line Profile Analysis (XRDLPA) systematically as a function of temperature and time to understand the evolution of microstructure in the very early stages of annealing.

## 2. EXPERIMENTAL

The chemical composition of alloy D9 is given in Table 1. The steel was obtained in the form of rods of 30mm diameter in the hot rolled condition. Rods of 26mm diameter were machined from these hot rolled rods and given a solution annealing treatment at 1373 K for 1/2 h followed by water quenching.

The highly deformed D9 powder was prepared from these annealed D9 rods with the help of finely threaded jewellery files. Fine particles in the size range 74-88μm were extracted from this D9 powder using 170 and 200 mesh sieves.

The High Temperature X-ray Diffraction experiments were carried out on these samples using the Bruker AXS D8 Advance Diffractometer with the Anton Paar high temperature attachment HTK 16. The attachment consisted of a Platinum (Pt) strip heater on which the sample is mounted. The whole stage was isolated from the surroundings by a vacuum chamber which was connected to a turbo molecular pump to maintain a high vacuum. During the experiments, the vacuum inside the chamber was better than $5 \times 10^{-5}$ mbar.

A required amount of the D9 powder sample was collected on a slide and then mixed with a few drops of isopropyl alcohol and a drop of Zapon$^{TM}$ which acted as a binder. This mixture was then pasted uniformly on the Pt strip heater in a rectangular shape. Special care was taken during the mounting of the sample to maintain the uniformity in the dimension and packing, for all the experiments.

Two different experiments using high temperature X-ray Diffractometer were carried out on the samples using Co-K$_\alpha$ radiation. In one of the experiments, an attempt was made to study the kinetics of the evolution of the microstructure of the samples at different elevated temperatures 673K, 773K, 823K, 838K, 845K, 853K and 873K. Since, the rearrangement of dislocations is a thermally assisted phenomenon, we had to restrict the annealing studies up to 873K so that the related kinetics of the microstructural evolution at a very early stage could be detected by XRD. In each case the sample was brought to the desired temperature at a rate of 1.66 K/sec and then the X-ray scan was started. To understand the kinetics of the microstructural evolution, it is required to collect the information of the evolution *in situ* at a close interval of time. Since the time required for a full profile scan was much larger (almost around 30 minutes with the scan step of 0.02º taken at 0.5 sec/step) the study was limited to the highest intensity peak (111) only which is also the slip plane of the Ti-modified austenitic stainless steel. The scan was performed between 48º and 53.5º which required a minimum scan time of 222 seconds including the instrument adjustment time after optimizing different instrumental parameters for obtaining an acceptable peak to background ratio for the (111) peak. One hundred and forty five number of X-ray scans for (111) peak were collected successively within a time of 9 hours to understand the kinetics.

In the other experiment, the X-ray data was collected *in situ* at room temperature and subsequently at different elevated temperatures ranging from 323K to 873K. At each temperature, the scan range of 2θ was 26° to 122° with a step of 0.02° taken at 0.5 sec/step. Each sample was heated to the required set temperature and a soaking time of 9 hours was allowed for the sample to stabilise its microstructural evolution before acquisition of the data. The experiment was performed several times to check the reliability and repeatability of the observed data.

TEM studies were carried out on the deformed and annealed (annealed at 873K for 9 hours) powder samples. In each case, a small quantity of the powder was suspended in acetone solution and the solution was sonicated for 7 hours. A drop of this solution was dropped onto TEM grids and was allowed to dry and stored in a vacuum dessicator. The micrographs were obtained using a JEOL 200kV TEM at the unit of Nanoscience and Technology, IACS, Kolkata, India.

## 3. METHOD OF ANALYSIS

### 3.1. *Single Peak Analysis:*

The single peak analysis of the diffraction pattern of the (111) plane has been performed to characterise the *in situ* time dependent evolution of the microstructure at different elevated temperatures. The true integral breadth ( $\beta$ ) of each XRD (111) peak profile was evaluated after correcting for the Debye Waller factor at that particular temperature and also for the instrumental broadening contribution. This $\beta$ was used to calculate the volume weighted average domain size ($D_v$) from the Scherrer's equation [27]:

$$D_v = \frac{0.9\lambda}{\beta\cos\theta} \tag{1}$$

### 3.2. Modified Rietveld Method

The microstructural variation with temperature has been studied by full powder pattern fitting technique using the Modified Rietveld method on the diffraction profiles. In this method, the diffraction profiles have been modelled by the pseudo-Voigt function using the program LS1 [28]. This program includes the simultaneous refinement of the crystal structure and the microstructural parameters like the domain size and the microstrain within the domain. The method involves the Fourier analysis of the broadened peaks. Considering an isotropic model, the lattice parameter ($a$), surface weighted average domain size ($D_s$) and the average microstrain $<\varepsilon_L^2>^{1/2}$ were used simultaneously as the fitting parameters to obtain the best fit. Having obtained the values of $D_s$ and $<\varepsilon_L^2>^{1/2}$, the average domain size and microstrain were further refined using anisotropic model to estimate the effective domain size ($D_e$) for each crystallographic plane.

## 4. RESULTS AND DISCUSSION

The variation of $\beta$ with time is obtained from the single peak analysis at different temperatures. The value of $\beta$ becomes almost stationary at different temperatures after 9 hours. Hence it can be assumed that there is saturation in the evolution of the microstructure which helped us to choose a definite soaking time at each temperature for the second experiment.

Fig 1 represents the XRD profile of the samples at room temperature and at higher temperatures after soaking of 9 hours. The figure clearly reveals the changes in the intensity patterns due to the change in the temperature. The inset of Fig 1 shows changes in the intensity of the (111) peak in an expanded scale.

Fig 2 represents the Rietveld fit of a typical XRD profile at a temperature of 823K. The variations of $D_s$ and $<\varepsilon_L^2>^{1/2}$ of these samples are shown as a function of temperature in Fig 3 and Fig 4 respectively.

Significant changes were observed in the values of $D_s$ with temperature. The average size of the domain was found to be around 14nm at room temperature. It is seen that initially the increase in $D_s$ is quite small up to the temperature around 673K but a steep rise is observed after this temperature, reaching a value of ~ 29nm at 873K. The formation of domains at room temperature and at high temperature depends on the arrangement of dislocations during the process of deformation and their mobility by glide and climb during annealing [29]. The steep increase in $D_s$ after 673K can be attributed to the higher probability of dislocation rearrangement and annihilation at high temperature by the process of climb which provides an additional degree of freedom for the movement of dislocations.

The variation of the values of the effective domain size $D_e$ at different crystallographic planes shows strong anisotropy at all temperatures (Fig. 5). The increase in $D_e$ with temperature up to 673K may be attributed to the glide process. The movements of dislocations in the crystallographic planes (111) and (220) occur by easy glide and hence the variation in $D_e$ with temperature is monotonic. On the contrary, the movement of dislocations by glide is restricted for the crystallographic planes (200) and (311) and an anomalous behaviour of $D_e$ with temperature up to 673K is observed. However at high temperature, the climb process becomes active which minimises this anomaly.

On the other hand, the microstrain values $<\varepsilon_L^2>^{1/2}$ (Fig. 4) did not change significantly up to the temperature of 873K. Hence it can be conjectured that the change in peak profiles is primarily governed by the variation in the domain size. In our next part of the discussion, the variation in the peak broadening with temperature and time has been considered to arise solely from the domain size.

In order to study the kinetics of the growth of the domains, the $\beta$ of the (111) peak estimated from our first experiment were used to obtain the evolution of normalised $D_v$ (with respect to $D_0$ at each data set, where $D_0$ is the initial domain size) with time at different temperatures, as shown in Fig 6. It is clearly evident from the figure that the growth rate of domains at the initial stage is significantly high as compared to that of the later stage. In a heavily deformed sample, the high population of loosely bound dislocations introduced into the sample form tangles as shown schematically in Fig. 7(a). When thermal energy is supplied to the sample, these dislocations rearrange themselves as shown in Fig. 7(b) resulting in a decrease in the configurational entropy of the system. Consequently the internal energy is also lowered, which in effect, reduces the total free energy (Helmholtz free energy) of the system. Thus the overall dynamical process leads to the formation of domains as shown in Fig. 7(c, d). Due to this rearrangement of dislocations, the localized density of dislocations at the domain boundary becomes very high as compared to the interior of the domain. As a result, the probability of annihilation of these dislocations increases leading to the growth of the domains as shown in Fig. 7(e). Thus it can be conjectured that the evolution of domain growth in the very early stages of annealing as seen in Fig. 6, is governed by two simultaneous mechanisms, one is rearrangement of dislocations which is predominant in the initial stage (stage I), and the other is their annihilation which is dominant in the later stage (stage II). We have not considered the dynamics of domain rotations in our discussion as it is well known that this plays a significant role at high temperature [10].

In the initial stage, the loosely bound dislocations participate in the formation of domains and this process continues till a low angle tilt boundary with high dislocation density is achieved and the formation of domains is almost complete. Consequently, the rate of increase in domain size is significantly higher in stage I as compared to stage II (Fig. 6). In stage II, this rate decreases drastically as the domains almost attain a fixed size and the entangled dislocations at the domain boundary annihilate each other.

These two distinctly different dynamical processes responsible for the evolution of the domains in the very early stage of annealing can be mathematically modelled as follows:

Had the process of rearrangement of dislocations been the sole mechanism for the increase in the size of the domains, these would certainly lead to a saturated average domain size ($D_m$). Hence, at any instance of time for a fixed temperature, the rate of increase of domain size ($D_v$) will be proportional to the difference ($D_m - D_v$), i.e.

$$\frac{dD_v}{dt} = k_1(D_m - D_v)$$

where $k_1$ is the rate constant. This leads to

$$D_v(t) = D_m - (D_m - D_v|_{t=0}) \exp(-t/\tau) \qquad (2)$$

where $\tau = 1/k_1$, which is the characteristic time of the dynamical process.

In the dynamical process involving annihilation of dislocations in stage II, the annihilation rate will be high for small sized domains as the average density of dislocations per unit volume is more for small domains, resulting in higher probability of annihilation. Hence the rate of increase of the domain size can be mathematically modelled as:

$$\frac{dD_v}{dt} = \frac{k_2}{D_v^n}$$

where $k_2$ is the rate constant. This leads to the equation

$$D_v^2 = D_0^2 + Bt \qquad (3)$$

with $n = 1$. Here, $D_0$ is the initial average domain size of this process and $B$ is a constant.

Equations (2) and (3) have been used to model the increase in the domain size as seen in Fig. 6. The solid and the dotted lines represent the final fit of the experimental data using equations (2) and (3) respectively. It is seen clearly from Fig. 6 that the first mechanism (rearrangement of dislocations) switches over to the second mechanism (annihilation of dislocation) at a lower time with the increase in temperature. This can be explained by the thermally activated climb process which becomes easier at high temperature; thus facilitating the rearrangement and annihilation of the dislocations to take place much faster. However, below 823K it is interesting to note that the data at temperatures 673K and 773K could be fitted with equation (2) only. This signifies that the small increase in the domain size results due to the rearrangement of the dislocations only. Hence it can be conclusively stated that below 823K, even if the annealing is done for a sufficiently long time, complete recovery would never be achieved in Ti- modified austenitic stainless steel.

TEM studies were carried out at room temperature on heavily deformed powder sample and on annealed sample at 873K for 9 hours and the micrographs are shown in Fig. 8(a) and 8(b) respectively. Fig. 8(a) shows the presence of highly strained regions due to the formation of high dislocation tangles throughout the matrix as expected in a heavily deformed structure. Annealing at 873K for 9 hrs still did not result in significant removal of these structures which is clearly seen by the presence of strained region even at 873K in Fig. 8(b). It is also seen that the spacing between the consecutive distorted regions is of the order of 10nm for heavily deformed sample and 20 nm for the annealed sample. These values correspond to the coherent region (domain) as detected by the XRD.

## 5. CONCLUSION

The microstructural evolution of heavily deformed D9 powder samples during the very early stage of annealing has been studied *in situ* with time and temperature by high temperature X-ray diffraction techniques. A systematic increase in the domain size was observed with the increase in temperature. The kinetics of the microstructural evolution has been found to follow mainly two different mechanisms, one is the rearrangement of dislocations and another is their annihilation. The climb process becomes active at high temperature which finally governs the kinetics of both rearrangement and annihilation of dislocations.

## 6. ACKNOWLEDGEMENT

One of the authors (A. D.) acknowledges CSIR, India for providing financial support.

**TABLE 1:** Chemical Composition in wt% of Alloy D9

| C | Mn | Ni | Cr | Mo | N | Ti | S | P | Fe |
|---|----|----|----|----|----|----|----|----|----|
| 0.05 | 1.50 | 15.04 | 15.09 | 2.26 | 0.006 | 0.21 | 0.003 | 0.01 | Balance |

FIGURE CAPTIONS:

1. XRD profiles of heavily deformed D9 powder sample at different temperatures.

2. Rietveld fit of a typical XRD profile at a temperature of 823 K.

3. Variation of surface weighted average domain size $D_s$ for the D9 powder sample as a function of temperature.

4. Variation of average microstrain $<\varepsilon_L^2>^{1/2}$ in the D9 powder sample as a function of temperature.

5. Variation of effective domain size $D_e$ with temperature along different crystallographic planes.

6. Evolution of normalized volume weighted domain size $D_v$ with time at different temperatures for the D9 powder sample. The solid and the dotted lines represent the final fit of the experimental data using equations (2) and (3) respectively.

7. Schematic representation of the (a) randomly distributed dislocations in heavily deformed sample, (b) initial stage of dislocation rearrangement, (c) formation of domains with rearrangement of dislocation, (d) annihilation of dislocation and (e) growth of domains due to dislocation annihilation.

8. Transmission Electron Micrographs for (a) heavily deformed D9 powder sample, (b) sample after 9 hour annealing at a temperature of 873K.

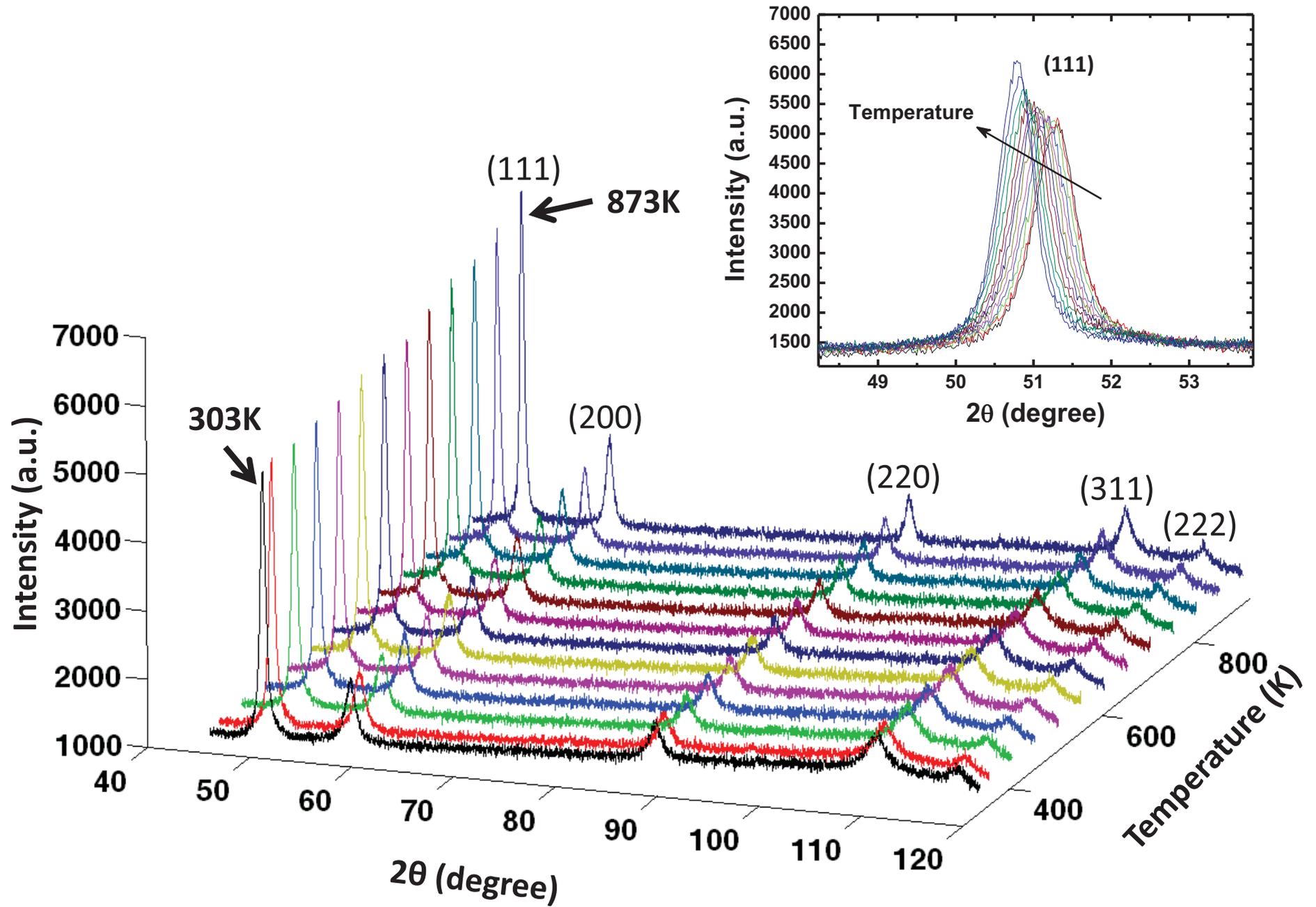

Fig. 2

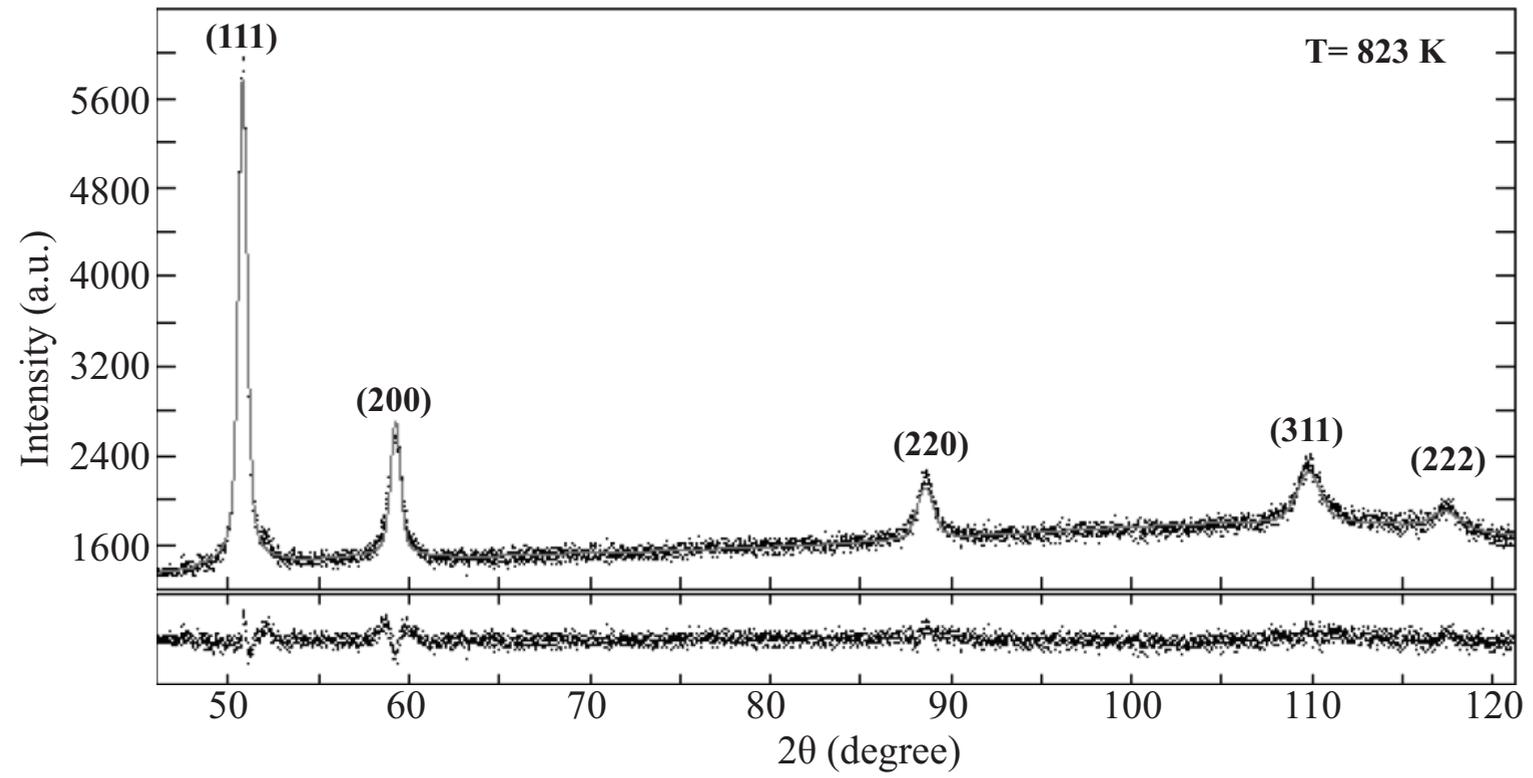

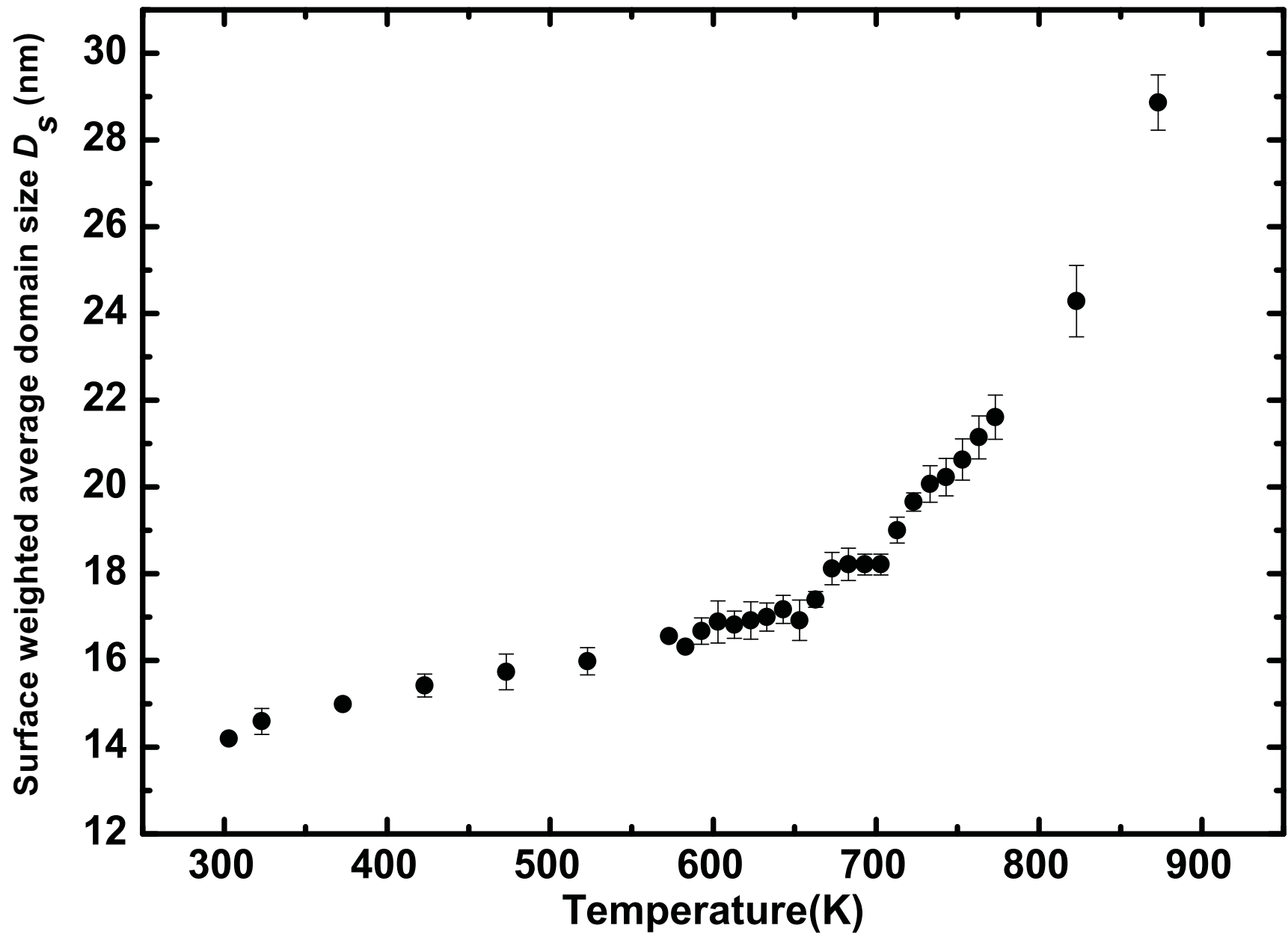

Fig. 4

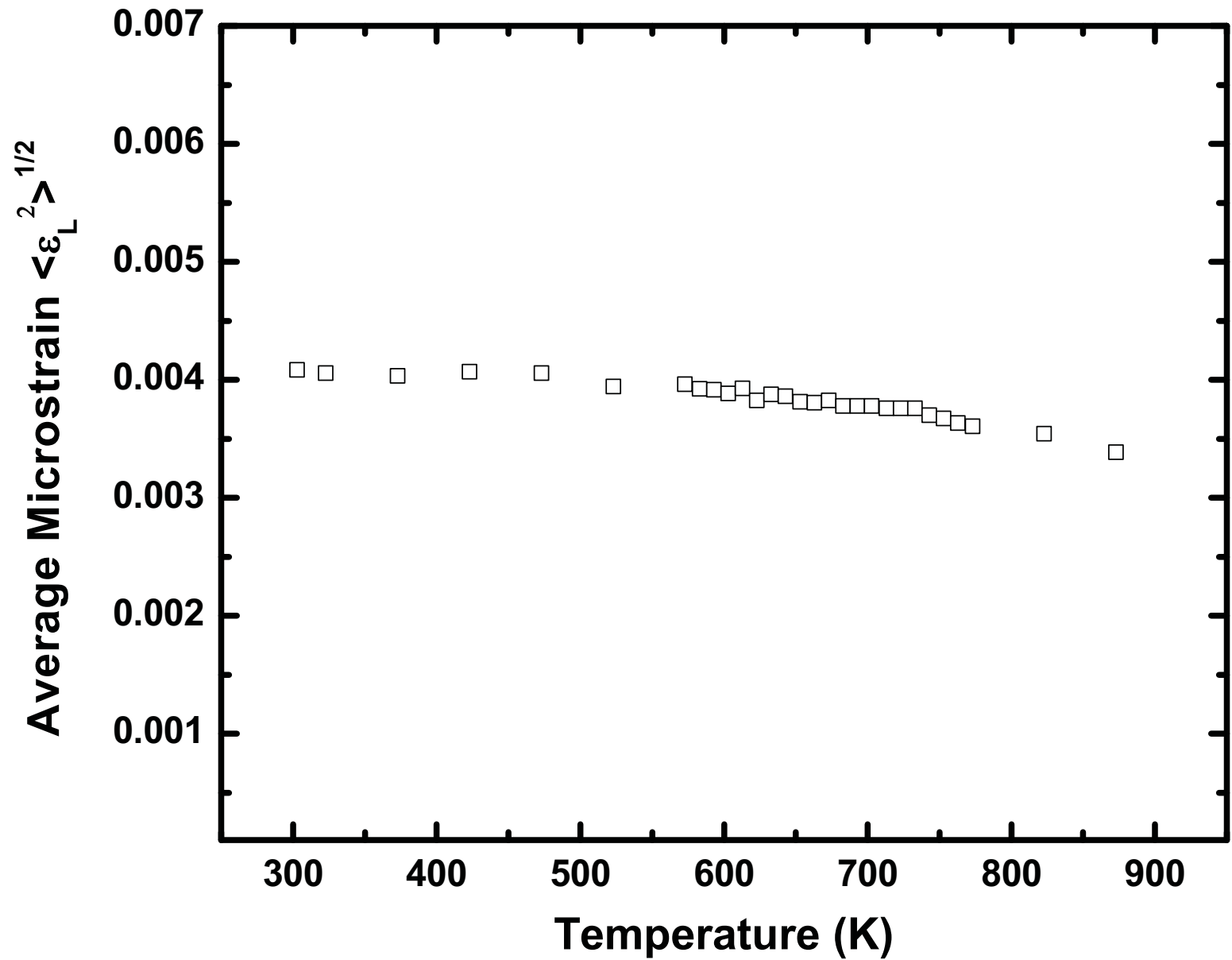

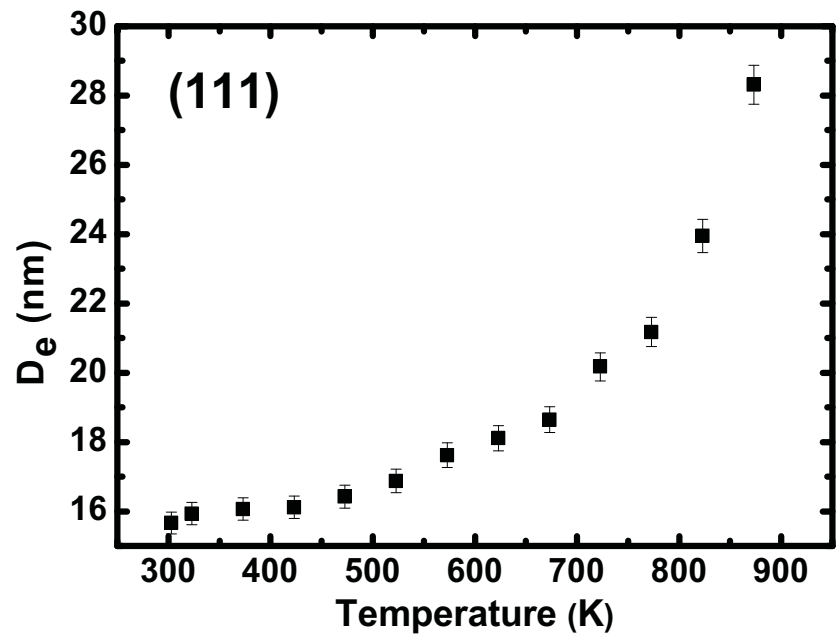

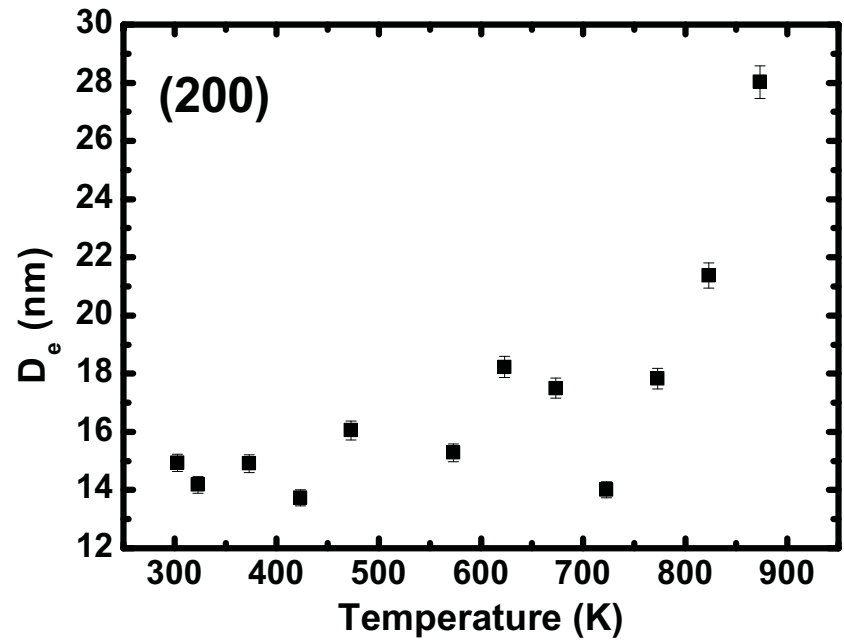

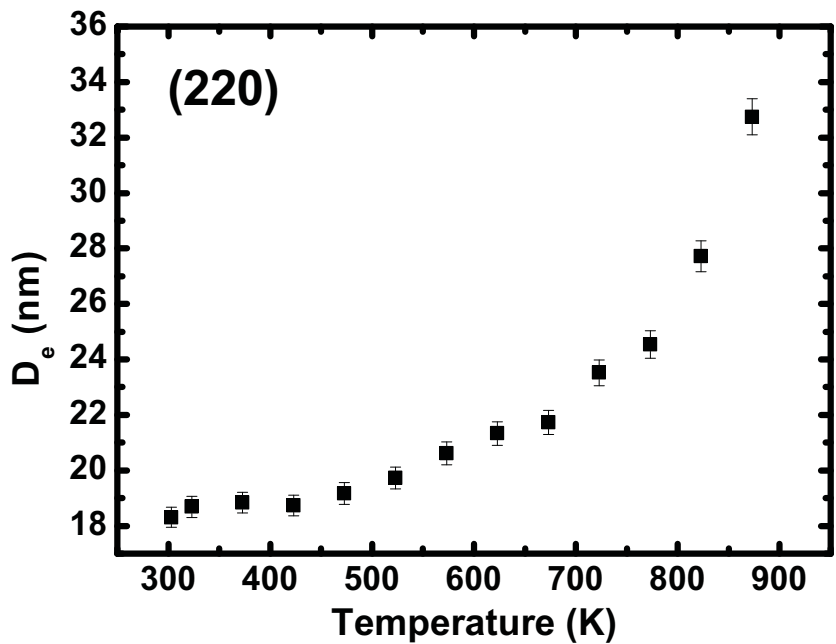

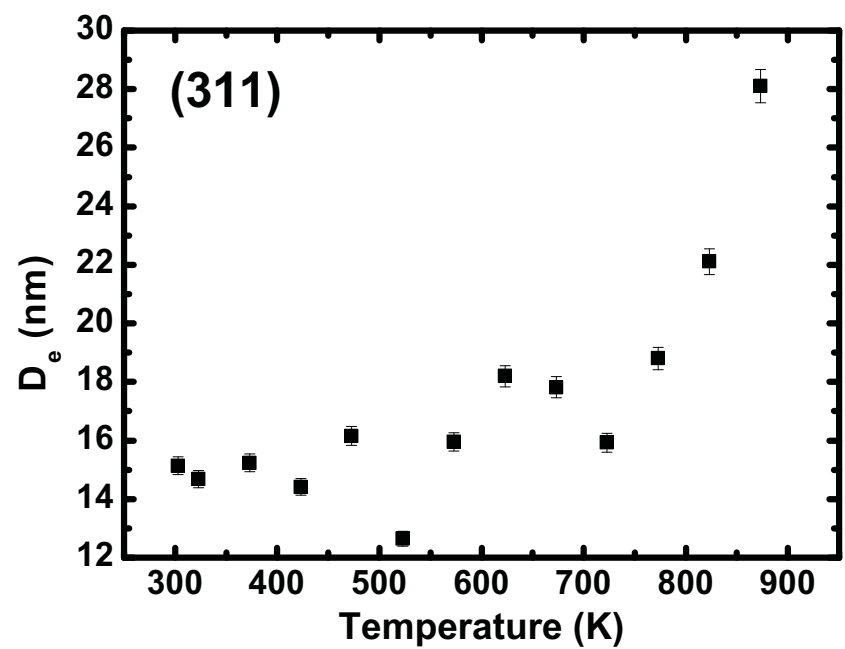

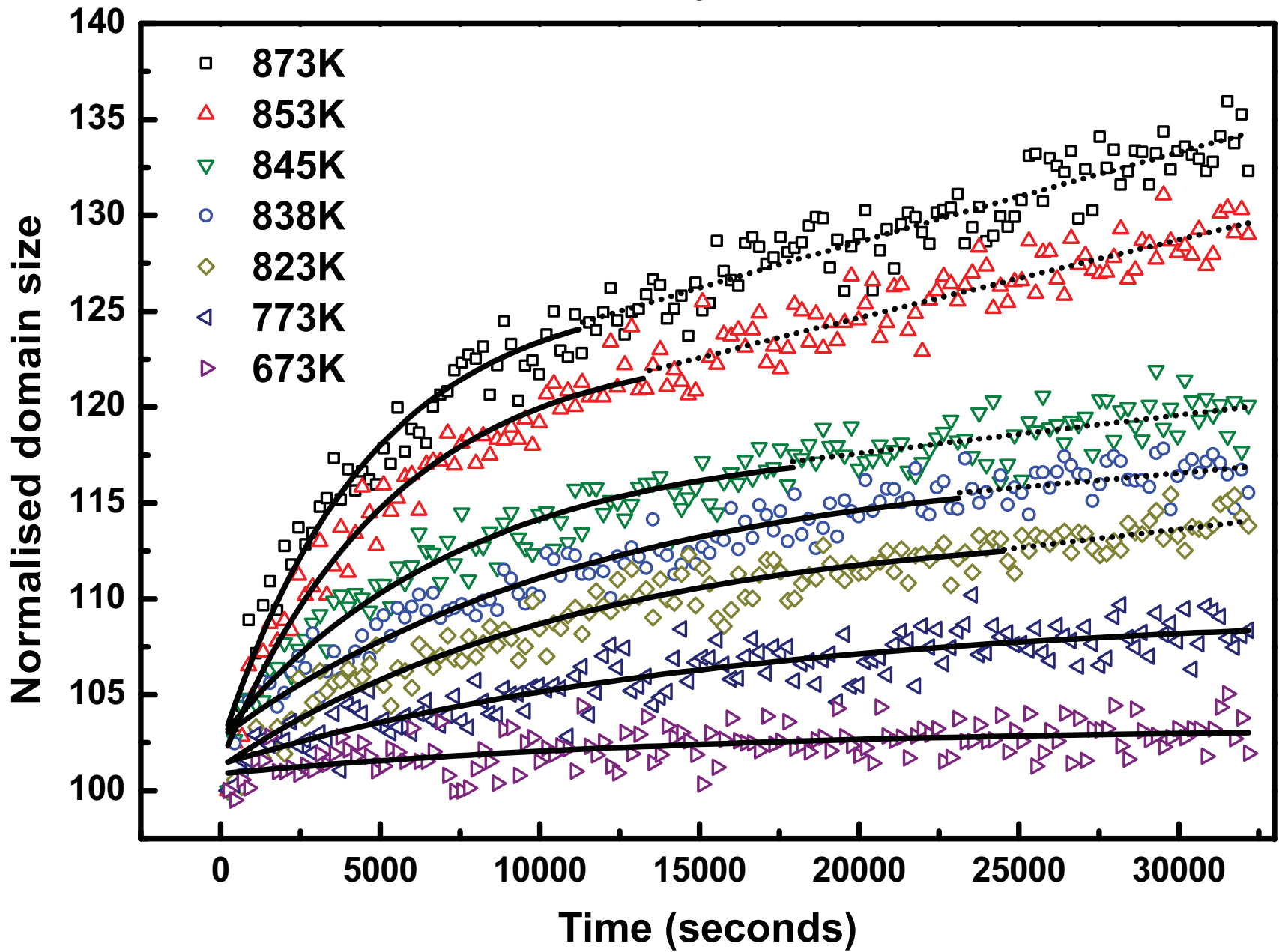

Fig. 6

Fig. 7

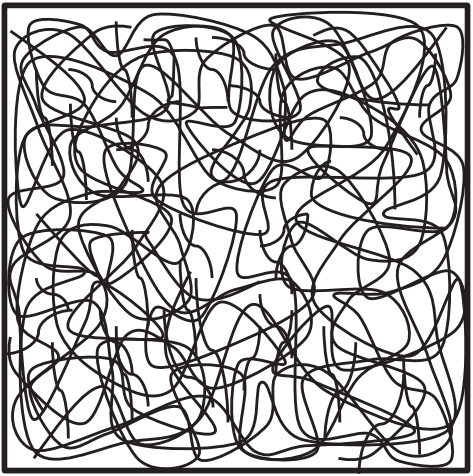

(a)

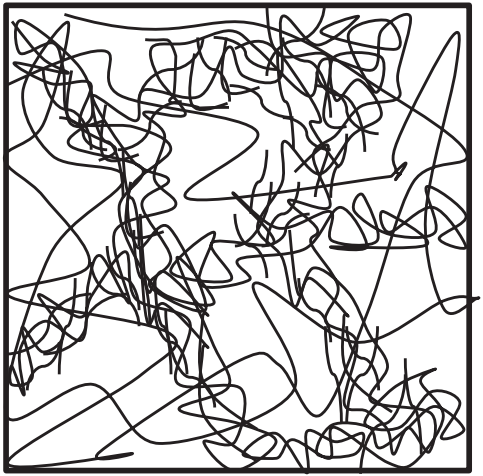

(b)

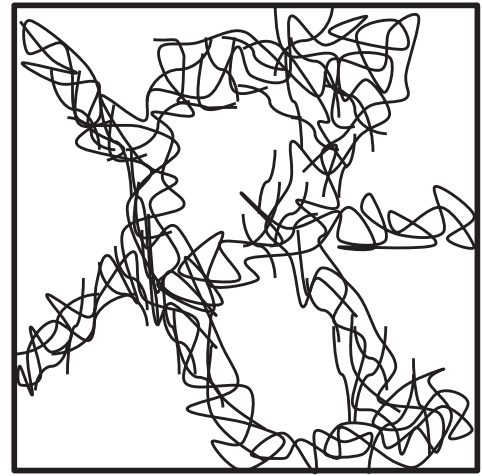

(c)

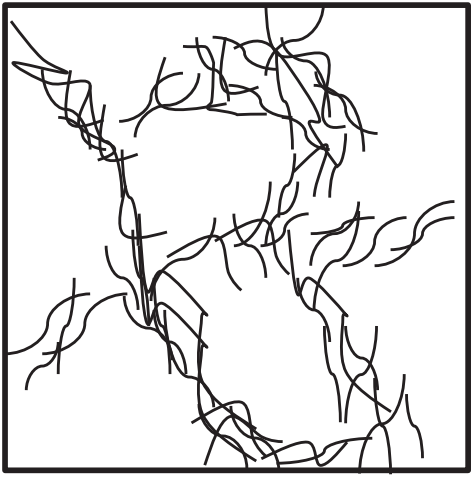

(d)

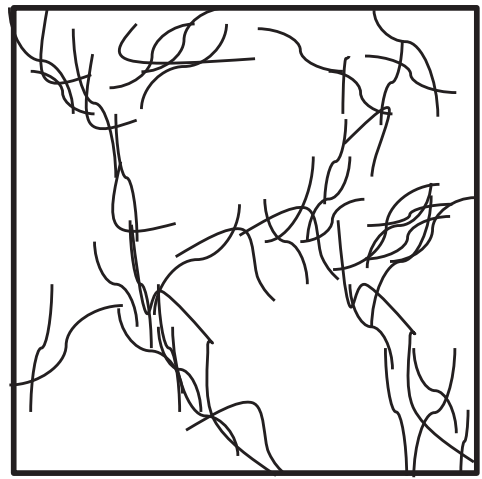

(e)

Fig. 8

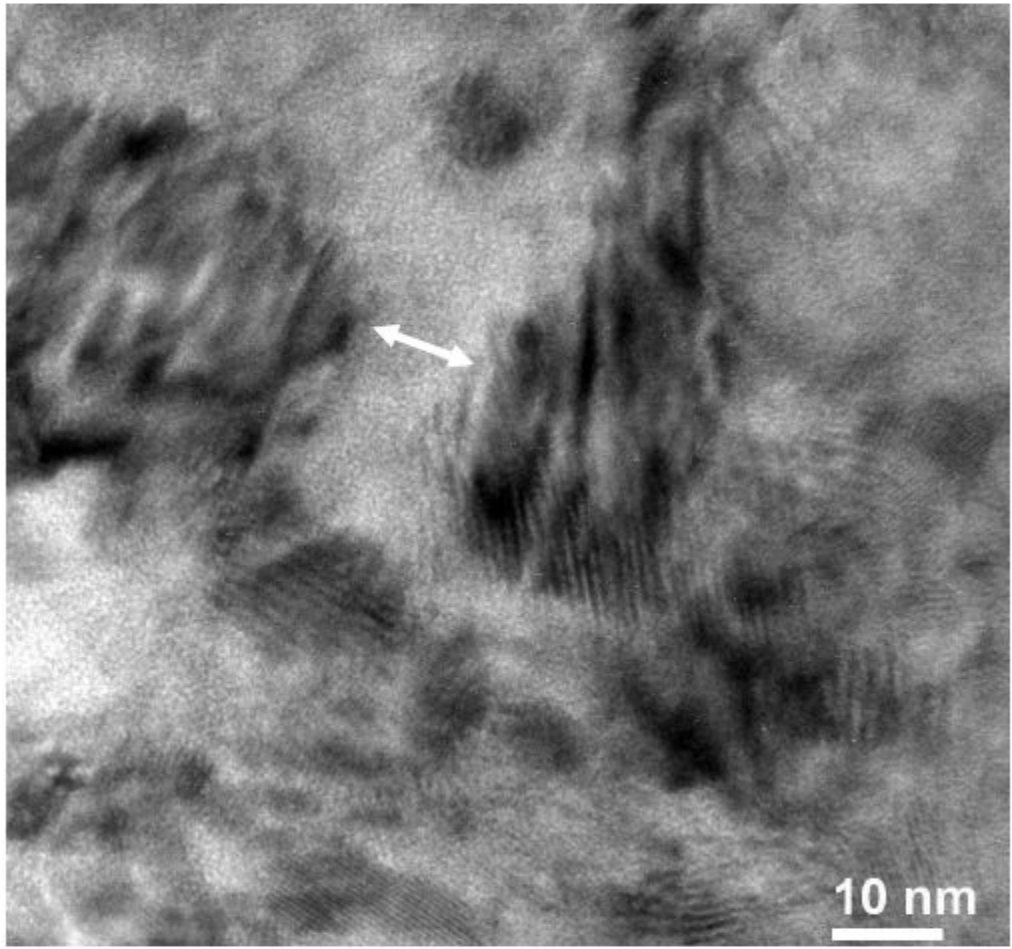

**(a)**

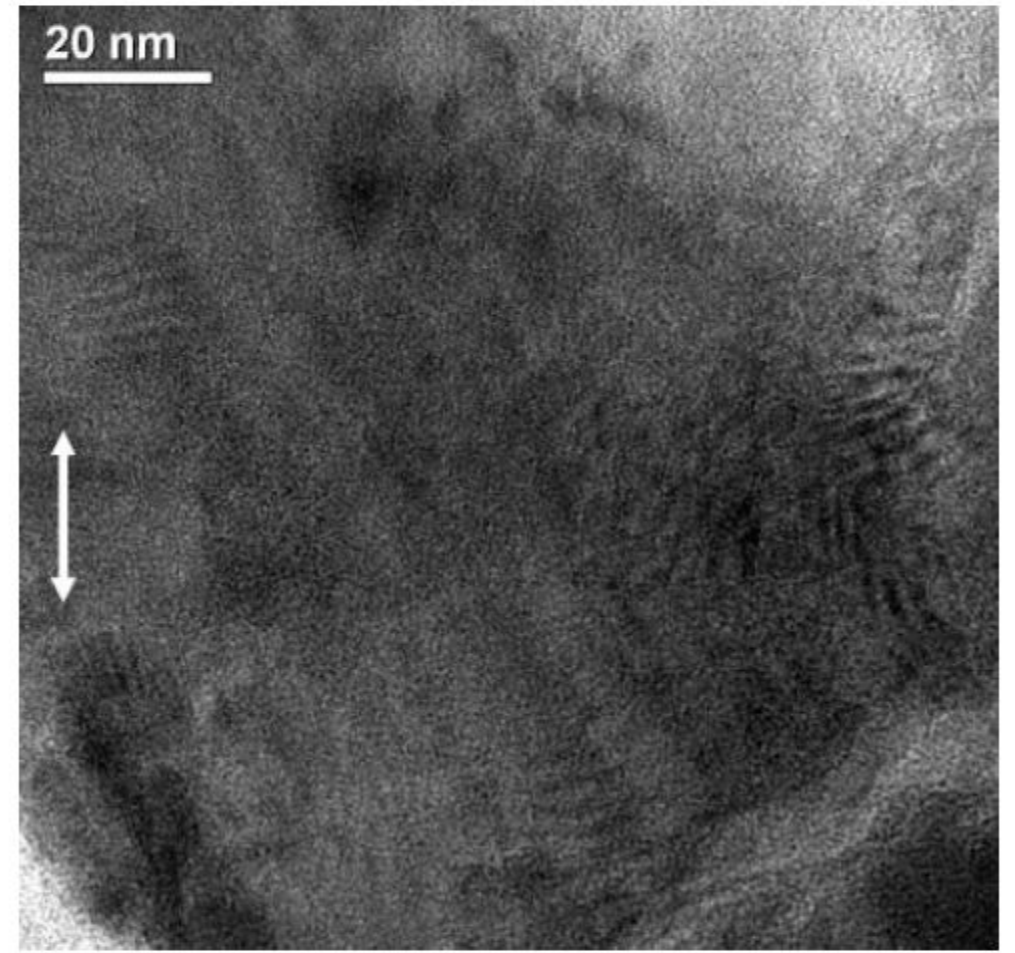

**(b)**